\documentclass[english,manuscript]{aastex}
\usepackage[T1]{fontenc}
\usepackage[latin9]{inputenc}
\usepackage{array}
\usepackage{float}
\usepackage{bm}
\usepackage{amstext}
\usepackage{graphicx}
\usepackage{esint}

\makeatletter

\usepackage{aas_macros}

\makeatother

\usepackage{babel}
\begin{document}

\title{Helioseismic inversion to infer depth profile of solar meridional
flow using spherical Born kernels }

\author{K. Mandal$^{1}$, S. M. Hanasoge$^{1}$, S. P. Rajaguru$^{2}$, H.
M. Antia$^{1}$}

\affil{$^{1}$Tata Institute of Fundamental Research, Mumbai, India}

\affil{$^{2}$Indian Institute of Astrophysics, Bengaluru, India}
\begin{abstract}
Accurate inference of solar meridional flow is of crucial importance
for the understanding of solar dynamo process. Wave travel times,
as measured on the surface, will change if the waves encounter perturbations
e.g. in the sound speed or flows, as they propagate through the solar
interior. Using functions called sensitivity kernels, we may image
the underlying anomalies that cause measured shifts in travel times.
The inference of large-scale structures e.g meridional circulation
requires computing sensitivity kernels in spherical geometry. \citet{2017ApJ...842...89M}
have computed such spherical kernels in the limit of the first-Born
approximation. In this work, we perform an inversion for meridional
circulation using travel-time measurements obtained from $6$ years
of SDO/HMI data and those sensitivity kernels. We enforce mass conservation
by inverting for a stream function. The number of free parameters
is reduced by projecting the solution on to cubic B-splines in radius
and derivatives of the Legendre-polynomial basis in latitude, thereby
improving the condition number of the inverse problem. We validate
our approach for synthetic observations before performing the actual
inversion. The inversion suggests a single-cell profile with the return-flow
occurring at depths below $0.78\,R_{\odot}$.
\end{abstract}

\section{Introduction}

Observation of solar meridional circulation is challenging because
of its small magnitude compared to other flows at the solar surface.
Improvements in observational techniques have made it feasible to
reliably infer the profile of meridional circulation at the surface
and near surface layers through a variety of techniques e.g., Doppler-shift
measurement \citep{1979SoPh...63....3D,1996Sci...272.1306H,2010ApJ...725..658U},
tracking of surface features e.g., small scale magnetic remnants \citep{1993SoPh..147..207K,2012ApJ...760...84H},
ring-diagram analysis \citep{1998ApJ...504L.131S,1999ApJ...512..458B,2002ApJ...570..855H,2008SoPh..252..235G,2010ApJ...717..488B},
time-distance helioseismology \citep{1997Natur.390...52G}. All these
studies show that the magnitude of meridional circulation at mid-latitudes
in the near-surface region is 10--20 m s$^{-1}$ at mid-latitudes,
depending on the phase of the solar cycle and is directed poleward
in both hemispheres. Mass conservation requires there to be an equatorward
return flow below the solar surface. The location of the return flow
plays a crucial role in some models of the solar dynamo. There have
been several attempts to understand the internal structure of meridional
circulation using helioseismic techniques e.g., global helioseismology
\citep{2013ApJ...778L..38S,2013SoPh..287..129W}, time-distance helioseismology
\citep{2013ApJ...774L..29Z,2015ApJ...805..133J,2015ApJ...813..114R,2017ApJ...845....2B,2017ApJ...849..144C}
etc. Such studies are very important to constrain dynamo models \citep{1995A&A...303L..29C,2010ApJ...722..774D}
since meridional circulation is believed to carry magnetic field as
well as transport energy and angular momentum in the convection zone.

These studies have arrived at a variety of conclusions. From the analysis
of two years of continuous data taken by SDO/HMI, \citet{2013ApJ...774L..29Z}
reported multiple cells in depth with a return flow starting at about
$0.9\:R_{\odot}$ and the second cell below $0.82\,R_{\odot}$. Using
2 years of GONG data, \citet{2015ApJ...805..133J} obtained results
that agreed with \citet{2013ApJ...774L..29Z} about the shallow return
flow but could not find the second cell in the deeper convection zone.
Recently, \citet{2017ApJ...849..144C} using $7$ years of SDO/HMI
data covering $2010$ May 1 to $2017$ April $30$ found qualitatively
similar profile as \citet{2013ApJ...774L..29Z}. These studies did
not apply the mass conservation constraint. Using 4 years of HMI data,
\citet{2015ApJ...813..114R} found a single-cell profile with return
flow below $0.77\,R_{\odot}$ using stream functions which automatically
ensures mass conservation. On the other hand, \citet{2013ApJ...778L..38S}
found multiple cells in both latitude and depth, which is very different
from those found using time-distance helioseismology.

Systematics and noise significantly affect results obtained below
$0.9\,R_{\odot}$ using time-distance helioseismology. All studies
have considered data obtained from different instruments covering
different periods of time. Sensitivity kernels computed using ray
theory were used by \citep{2013ApJ...774L..29Z,2015ApJ...805..133J,2015ApJ...813..114R,2017ApJ...849..144C}
to invert for flows. Ray kernels are sensitive only to perturbations
along ray path. If the length scale of the perturbation is smaller
than the acoustic wavelength, results using ray kernels may not be
reliable \citep{2001ApJ...561L.229B,2004ApJ...616.1261B}.

The first-Born approximation, an alternative to ray theory, does not
suffer from the above limitation. Recently, \citet{2017ApJ...842...89M}
computed travel-time sensitivity kernels in the Born limit in spherical
geometry \citep[two other alternative approaches were proposed independently by][]{2016ApJ...824...49B,2017A&A...600A..35G}.
Using Born kernels, \citet{2017ApJ...845....2B} inverted for meridional
circulation and found return flow at $0.9\:R_{\odot}$ using a SOLA
\citep{1994A&A...281..231P} inversion technique. They found that
both single and multiple-cell profiles are consistent with their measured
travel times.

In this work, we use a stream-function approach, which automatically
takes into account mass conservation and compute relevant sensitivity
kernels. The other advantage of using stream functions is that both
radial and latitudinal components of the meridional flow may be simultaneously
derived from it.

\section{Travel-time measurements}

\begin{figure}
\centering{}\includegraphics[scale=0.55]{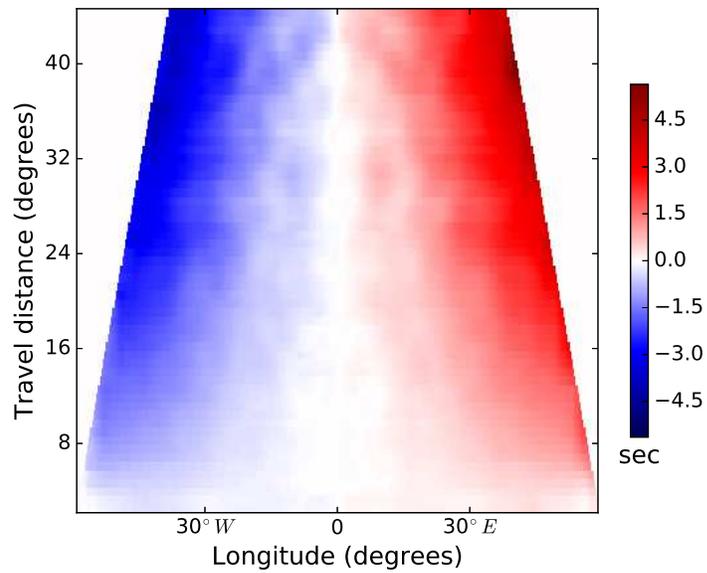}\caption{\label{fig:CTL}West minus east travel time, considered as center-to-limb
systematics are plotted. These has been subtracted from north minus
south travel time to obtain travel time shift due to solar meridional
flow which are shown in the left panel of Figure \ref{fig:data_error}. }
\end{figure}

\begin{figure}[t]
\begin{centering}
\includegraphics[scale=0.55]{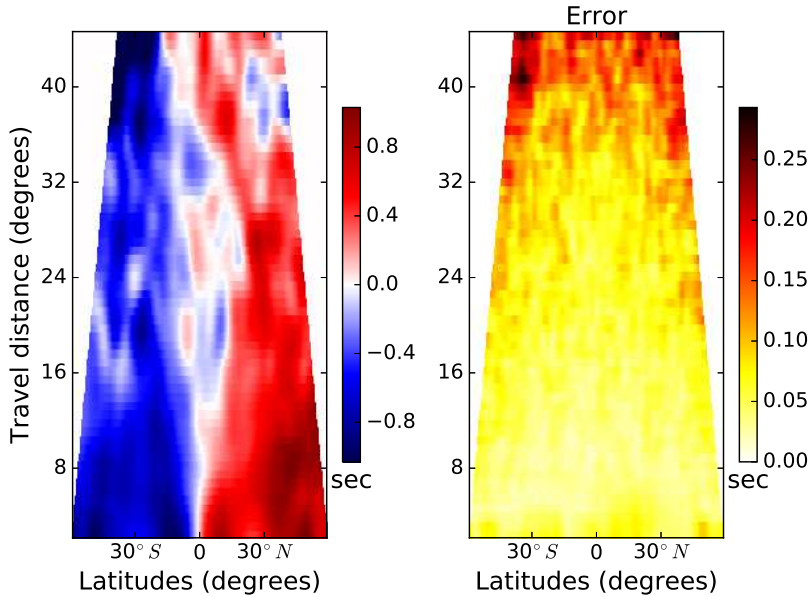}
\par\end{centering}

\caption{\label{fig:data_error}Left panel shows the measured travel times
for solar meridional flow and right panel shows corresponding errors
in the measurements. The travel time differences plotted are southward
minus northward travel times, i.e. $\delta\tau=\tau_{\text{NS}}-\tau_{\text{SN}}$,
where $\tau_{\text{NS}}$ is the travel time from north-arc to south-arc
and $\tau_{\text{SN}}$ is that in the opposite direction.}
\end{figure}

Helioseismic travel times capturing the signals due to meridional
flows are estimated the same way as described in detail in \citet{2015ApJ...813..114R}.
The basic data used are the full-disk Doppler observations made by
the Helioseismic and Magnetic Imager (HMI) onboard the Solar Dynamics
Observatory (SDO), and we have added two more years of data to that
of \citet{2015ApJ...813..114R} covering a total length of six years
($2010$ May $1$ through $2016$ April $30$). Each estimate of travel
time is made from a Gabor wavelet \citep{1997ASSL..225..241K} fit
to the monthly average of time-distance correlations computed in point-to-point
deep-focus arc geometry employing 60 travel distances. Travel distances
ranging between $2.16^{\circ}$ and $44.64^{\circ}$in steps of $0.72^{\circ}$,
covering depths from near the surface down to about $0.7\:R_{\odot}$.
A major uncertainty in travel time estimates is the center-to-limb
systematics - CLS \citep{2012ApJ...749L...5Z,2012ApJ...760L...1B,2015ApJ...813..114R,2013ApJ...774L..29Z}
whose magnitude is several times of the signal due to meridional flows.
We correct the N-S (or the meridional direction) travel times for
this systematics in the same way as originally proposed by \citet{2012ApJ...749L...5Z}
and also as done by \citet{2015ApJ...813..114R}: CLS is estimated
from the antisymmetric component of W-E travel times (Figure \ref{fig:CTL})
(the symmetric part corresponds to the solar rotation), which are
then subtracted from the N-S travel times. These W-E travel times
are estimated from the central W-E equatorial belt spanning 15 degrees
on either side of the equator. Recently, \citet{2017ApJ...849..144C}
have proposed a new empirical method to remove the systematics. They
measure travel-time shifts between two points located on the solar
disk for many different azimuthal angles and skip distances, followed
by solving a system of linear equations containing both center-to-limb
systematics and travel-time shifts due to meridional flow. In addition,
\citet{2017ApJ...849..144C} have removed data pixels containing fields
stronger than a threshold of $10$ G, following a method suggested
by \citet{2015ApJ...805..165L}. In our travel-time measurement process,
we have not removed the oscillation signals over the strongly magnetized
surface regions. It should be noted that surface magnetic fields have
been shown to corrupt the meridional flow measurements (\citealp{2015ApJ...805..165L});
although such influence of surface regions is not expected to majorly
affect the overall inferences on the deep structure of meridional
circulation inferred from 6 years of observations, a detailed analysis
with and without the surface magnetic regions' removal is necessary
to quantify the surface effects. However, detailed analyses of the
character of center-to-limb systematics and its dependence on the
frequency of acoustic waves (Rajaguru and Antia 2018, work in progress,
and \citealp{2018ApJ...853..161C}) indicate that further careful
study is needed to fully account for them in the current estimates
of meridional flow travel times. Measured travel times induced by
solar meridional flow from our analyses and corresponding errors have
been shown in Figure. (\ref{fig:data_error}). 

\section{forward modeling with spherical Born kernel}

An efficient approach to compute sensitivity kernels in spherical
geometry was proposed by \citet{2017ApJ...842...89M}. They showed
that Green's function can be expressed as \citep[Equation 12 and 13 from][]{2017ApJ...842...89M}
\begin{eqnarray}
G_{rr}(\mathbf{r},\mathbf{r}_{s},\omega) & = & \sum_{\ell}\frac{(2\ell+1)}{4\pi}\alpha_{\ell\omega}(r)P_{\ell}(\cos(\hat{\mathbf{r}}\cdot\hat{\mathbf{r}}_{s})),\label{eq:Greenfn2}\\
\mathbf{G}_{hr}(\mathbf{r},\mathbf{r}_{s},\omega) & = & \sum_{\ell}\frac{(2\ell+1)}{4\pi\omega^{2}\rho_{0}}\beta_{\ell\omega}(r)\bm{\nabla}_{h}P_{\ell}(\cos(\hat{\mathbf{r}}\cdot\hat{\mathbf{r}}_{s})),\label{eq:Greenfn1}
\end{eqnarray}
where $G_{rr}(\mathbf{r},\mathbf{r}_{s},\omega)$ and $\mathbf{G}_{hr}(\mathbf{r},\mathbf{r}_{s},\omega)=\left(G_{\theta r}\left(\mathbf{r},\mathbf{r}_{s},\omega\right),G_{\phi r}\left(\mathbf{r},\mathbf{r}_{s},\omega\right)\right)$
are radial and tangential components of the displacement of a wave
with temporal frequency $\omega,$ measured at $\mathbf{r}$ due to
a point source placed at $\mathbf{r}_{s}$, $P_{\ell}$ the Legendre
polynomial of degree $\ell$. The terms $\alpha_{\ell\omega},\,\beta_{\ell\omega}$
are obtained by solving a coupled differential equation \citep[Equation 10 in][]{2017ApJ...842...89M}
using a finite-difference based scheme for each harmonic degree $\ell$
and frequency $\omega$. We compute the pair $\left(\alpha_{\ell\omega},\,\beta_{\ell\omega}\right)$
once and use them to evaluate Green's function, which is used subsequently
in obtaining kernels. The expression of sensitivity kernels for stream
function, $K_{\psi}$, can be written in terms of Green's function
and its derivatives. The kernel computation time depends on the size
of the $(r,\theta,\phi)$ grid used in the analysis, maximum harmonic
degree for Green's function computation and resolution in frequency.
In this work, we consider $1398\times200\times400$ grid points spanning
$0.7\,R_{\odot}-1.0\,R_{\odot}$ in the radial direction and the entire
horizontal domain. We choose the harmonic degree in the range $20\leq\ell\leq100$
to compute Green's function using Equation (\ref{eq:Greenfn2}) and
(\ref{eq:Greenfn1}). We choose the frequency range $2.0$ mHz to
$4.5$ mHz, split uniformly into 1250 bins. It takes approximately
1.5 hours to compute a stream-function sensitivity kernel for one
source-receiver distance using 200 processors on a computer cluster.
It is indeed very expensive to compute sensitivity kernels for all
60 travel distances centered around 329 latitude points. Since we
do not consider line-of-sight effects in this work, we need to compute
sensitivity kernels only for 60 travel distances. We then translate
the kernels at different latitude points since these kernels depend
only on the source-receiver distance and not on their exact location.
We have validated these sensitivity kernels by choosing simple flow
profiles.

\section{inversion technique\label{sec:inversion-technique}}

Flows differently alter travel times of upstream and downstream propagating
waves. Sensitivity kernels for flows relate them to the travel time
difference, thus 
\begin{equation}
\delta\tau=\int_{\odot}K_{v_{r}}v_{r}d\mathbf{r}+\int_{\odot}K_{v_{\theta}}v_{\theta}d\mathbf{r}.\label{eq:dtau_vel}
\end{equation}
The contribution of the second integral is much larger than the first.
The rising and falling part of the $v_{r}$ component cancels out
whereas the contribution of two branches of the $v_{\theta}$ component
add to each other. Further, the magnitude of $v_{r}$ is also smaller.
Because of the small contribution from the $v_{r}$ component, it
is almost impossible to directly determine it from inversions. We
use the stream function, $\psi$, instead of velocity for inversion
which automatically takes into account mass conservation and both
components of velocity, $v_{r}$ and $v_{\theta}$ can be determined
from this single function using the following relations 
\begin{eqnarray}
\rho v_{r} & = & \frac{1}{r}\frac{\partial\psi}{\partial\theta}+\frac{\cos\theta}{r\sin\theta}\psi,\label{eq:streamfn1}\\
\rho v_{\theta} & = & -\frac{\partial\psi}{\partial r}-\frac{\psi}{r},\label{eq:streamfn}
\end{eqnarray}
where $\rho$ is the density of the medium. Substituting Equation
(\ref{eq:streamfn1}) and (\ref{eq:streamfn}) into Equation (\ref{eq:dtau_vel}),
we obtain 
\begin{equation}
\delta\tau=\int_{\odot}K_{\psi}\psi d\mathbf{r}\label{eq:kernel_psi}
\end{equation}
To compute sensitivity kernels for stream function $\psi$, we use
model S (\citealp{ChristensenDalsgaard:1996ap}) as the background solar
model \citep{2017ApJ...842...89M}. These kernels are highly sensitive
to surface layers and very weakly so to the base of the convection
zone, making it difficult to determine the profile of meridional circulation
at depth. Further, the density varies by 6 orders of magnitude over
the convection zone, while the velocity of meridional flow does not
vary to that extent. To account for this variation we invert for the
quantity $\psi^{\prime}=\psi/\rho$.

The size of the problem will become very large and ill-posed if we
perform inversions for $\psi$ at all spatial points. B-spline basis
function for both radial and latitude coordinates were used by e.g.
\citet{2015ApJ...813..114R} to represent the stream function. Recently,
\citet{2017arXiv170803464B} used B-spline basis functions successfully
for synthetic inversions of supergranules. This motivates us to use
B-spline functions to represent variations in the radial direction.
In the latitudinal direction, we use derivative of Legendre polynomial
for the inversion. The total number of parameters is reduced drastically
in this approach: 
\begin{equation}
\psi(r,\theta)=\sum_{i,\ell}a_{i\ell}B_{i}(r)\frac{dP_{\ell}(\cos\theta)}{d\theta},\label{eq:psi}
\end{equation}
where $a_{i\ell}$ are the coefficients of expansion which we need
to determine. The reason for choosing Legendre polynomials instead
of B-splines in the latitudinal direction is that fewer coefficients
are needed. In addition, we assume hemispheric symmetry of the meridional
circulation and consider only the derivative of even Legendre polynomials,
which reduces the number of parameters in the inversion and makes
the problem better posed. In all inversions, we aim to minimize the
misfit function, defined here as 
\begin{equation}
\chi_{\text{MF}}=\sum_{i}\left(\frac{\int K_{i}\psi(r,\theta)rdrd\theta-\delta\tau_{i}}{\sigma_{i}}\right)^{2}+\text{Regularization},\label{eq:misfit}
\end{equation}
where $i$ indicates a pair of observation points. The terms $K_{i}$
and $\delta\tau_{i}$ denote the corresponding kernel and travel-time
measurement. Despite the reduction in parameter space, we still need
to apply regularization because travel-time measurements are subject
to systematic and realization noise, which may be amplified in the
inversion. We use second derivative smoothing both in $r$ and $\theta$
\begin{equation}
\text{Regularization}=\int\left[\lambda_{r}\left(\frac{\partial^{2}\psi(r,\theta)}{\partial r^{2}}\right)^{2}+\lambda_{\theta}\left(\frac{\partial^{2}\psi(r,\theta)}{\partial\theta^{2}}\right)^{2}\right]r^{2}\sin\theta d\theta dr,\label{eq:Reg}
\end{equation}
where $\lambda_{r}$ and $\lambda_{\theta}$ are regularization parameters
which may be tuned to obtain a smooth solution. We apply the constraint
$\psi=0$ at the upper boundary. We use two different methods based
on the regularized least squares (RLS) technique to obtain unknown
coefficients in the Equation (\ref{eq:psi}) by minimizing the misfit
function.

\subsection{First method}

After substituting Equation (\ref{eq:psi}) into Equation (\ref{eq:misfit})
and taking derivatives with respect to unknown coefficient $a_{i\ell}$,
we obtain a system of equations which is written in matrix form, 
\begin{equation}
(M^{T}\Lambda M+M_{\text{Reg}})a=M^{T}\Lambda\tau,\label{eq:matrix_form}
\end{equation}
where 
\begin{equation}
M_{ij}=\int K_{i}(r,\theta)B_{k}(r)\frac{dP_{\ell}(\cos\theta)}{d\theta}rdrd\theta,\label{eq:M_matrix}
\end{equation}
\begin{equation}
\Lambda=\left(\begin{array}{cccc}
\frac{1}{\sigma_{1}^{2}} & 0 & 0 & ..\\
0 & \frac{1}{\sigma_{2}^{2}} & 0 & ..\\
.. & .. & .. & ..\\
.. & .. & 0 & \frac{1}{\sigma_{n}^{2}}
\end{array}\right),\label{eq:Lamda_matrix}
\end{equation}
$M_{\text{Reg}}$ is the Regularization matrix and it is obtained
from Equation (\ref{eq:Reg}). Each pair of $(k,\ell)$ determines
an unique index $j$ in Equation (\ref{eq:M_matrix}). The column
vector $a$ is composed of expansion coefficients $\{a_{i\ell}\}$
of Equation (\ref{eq:psi}) and is obtained by solving the Equation
(\ref{eq:matrix_form}). 
\begin{equation}
a=(M^{T}\Lambda M+M_{\text{Reg}})^{-1}M^{T}\Lambda\tau.\label{eq:inverse_eqn}
\end{equation}
The inverse of the matrix in Equation (\ref{eq:inverse_eqn}), which
may be obtained using many methods, is computed here using singular
value decomposition.

\subsection{Second method}

Misfit from Equation (\ref{eq:misfit}) with the regularization can
be satisfied if we solve following system of equations to obtain the
unknown coefficients in the Equation (\ref{eq:psi}) 
\begin{eqnarray}
\frac{1}{\sigma_{i}}\int_{\odot}K_{i}(r,\theta)\psi(r,\theta)\,r\,drd\theta & = & \frac{1}{\sigma_{i}}\delta\tau_{i},\label{eq:kernel_time_eqn}\\
\lambda_{r}\frac{\partial^{2}\psi(r_{k},\theta_{m})}{\partial r^{2}} & = & 0,\label{eq:norm_r}\\
\lambda_{\theta}\frac{\partial^{2}\psi(r_{k},\theta_{m})}{\partial\theta^{2}} & = & 0,\label{eq:norm_theta}
\end{eqnarray}
$\left(r_{k},\theta_{m}\right)$ is one of many points in the grid
on which we place constraints (\ref{eq:norm_r}) and (\ref{eq:norm_theta})
in order to obtain smooth solution. The Condition number of the matrix
for the second method is better than the first method.

\section{Inversion for synthetic data without and with noise\label{wo_noise}}

Systematic errors can appear from the fact that we use finite numbers
of Legendre polynomials and B-spline knots to expand the stream function.
This error decreases with increasing numbers of knots for B-spline
and degree of Legendre polynomials. However, the tradeoff is that
the condition number of the problem becomes correspondingly larger.
We define the misfit function for this case as 
\[
\chi_{\text{MF}}=\sum_{i}\left(\int K_{i}\psi(r,\theta)rdrd\theta-\delta\tau_{i}\right)^{2}+\text{Regularization}.
\]
In our work, we consider 40 knots in the radial direction and all
even values for $\ell$ with highest degree $16$ for Legendre polynomials
in the Equation (\ref{eq:psi}). In order to validate our inversion
algorithm, we first consider a few profiles e.g., single cell, double
cell and estimate the travel time corresponding to that profile by
forward modeling. We perform inversions with these travel-time measurements
and compare the retrieved profile with the original one.

After ensuring that systematic errors do not affect the inversion,
we add realization noise into the travel-time measurements (obtained
in the previous section). We assume that noise is Gaussian and randomly
perturb the travel time in proportion with errors found in observations.
We then perform inversions with these travel-time data and compare
the retrieved profile with the original one. In this case, we consider
the misfit function (\ref{eq:misfit}). We are only going to present
inversion results for noisy synthetic data in this paper. As expected,
inverted profiles obtained using noise-free measurements agree better
than those obtained using noisy travel times.

In order to get reliable noise estimates, we follow the approach of
\citet{2015ApJ...813..114R}. We randomly perturb the travel-time
according to the error in the observation. We estimate the velocity
and the associated standard deviation of these values, thereby propagating
measurement to inferential errors.

\begin{figure}[h]
\begin{centering}
\includegraphics[scale=0.55]{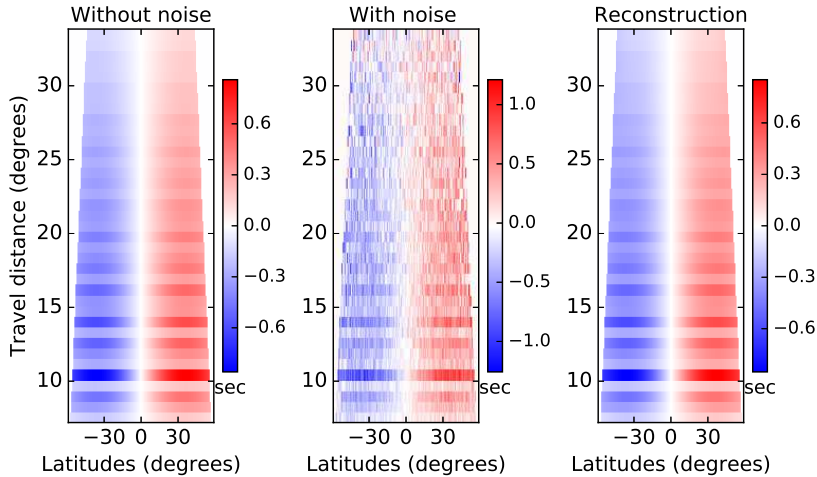} 
\par\end{centering}

\caption{\label{fig:synthetic_travel_time}Left panel shows travel times corresponding
to the single-cell profile (see panels a and c of Figure \ref{fig:synthetic_single}).
Mid panel: travel-times of left panel are randomly perturbed in proportion
with observed errors, later used for inversion are shown. Right panel:
travel-time obtained using the inverted profile of the stream function
using Equation (\ref{eq:kernel_psi}). }
\end{figure}

\begin{figure}[h]
\begin{centering}
\includegraphics[scale=0.6,clip=true]{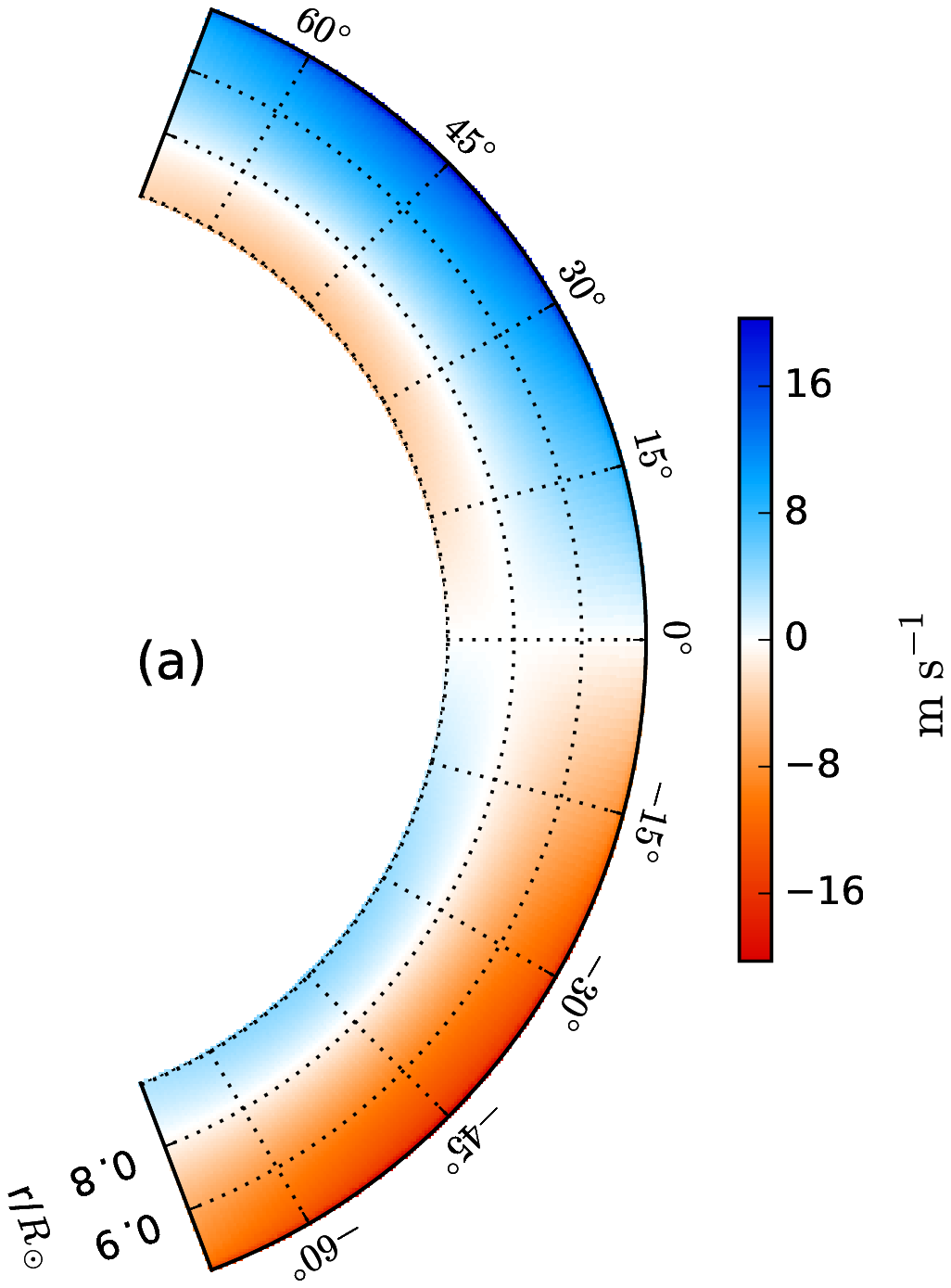}\includegraphics[scale=0.6,clip=true]{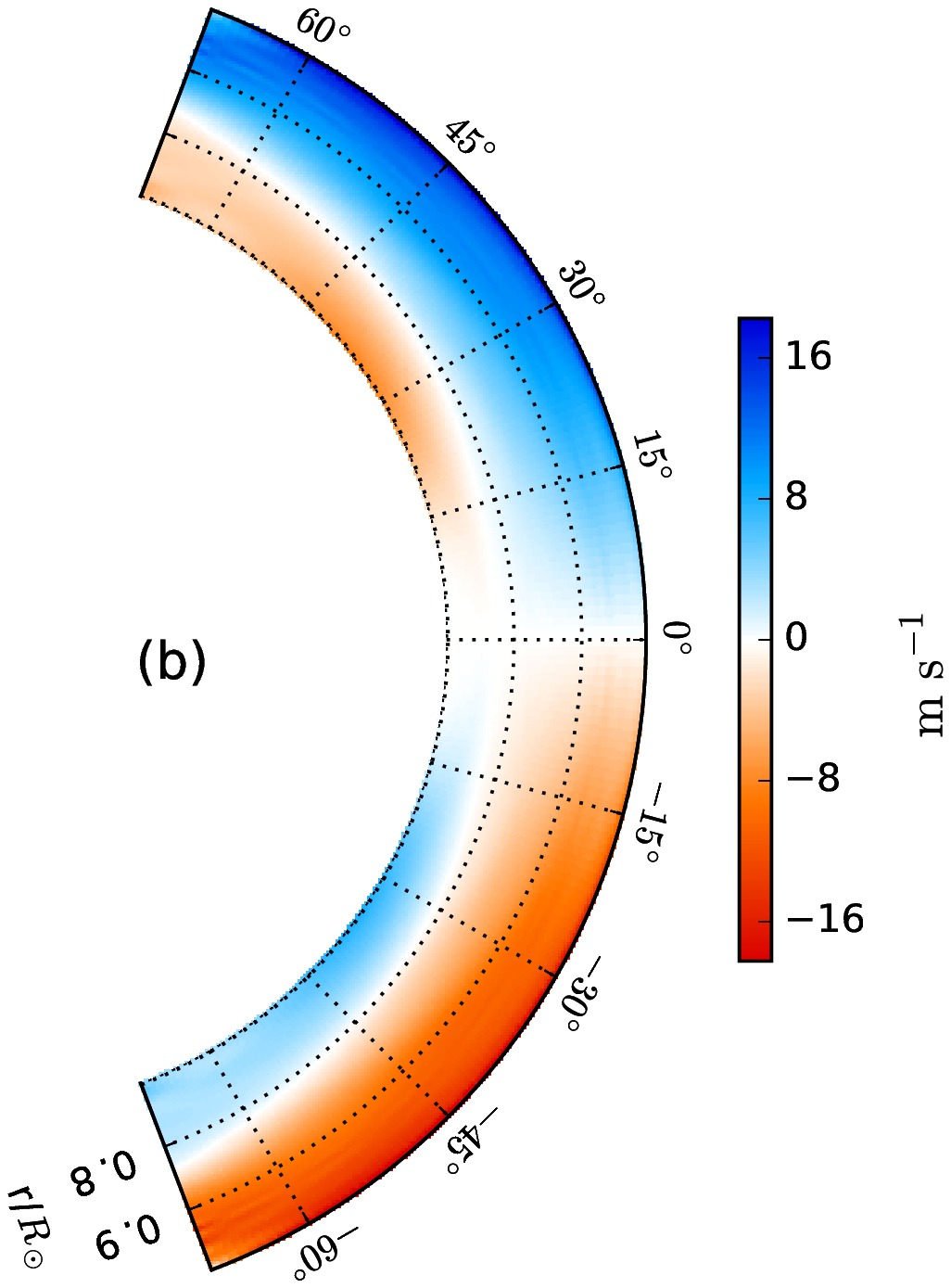} 
\par\end{centering}

\begin{centering}
\includegraphics[scale=0.6,clip=true]{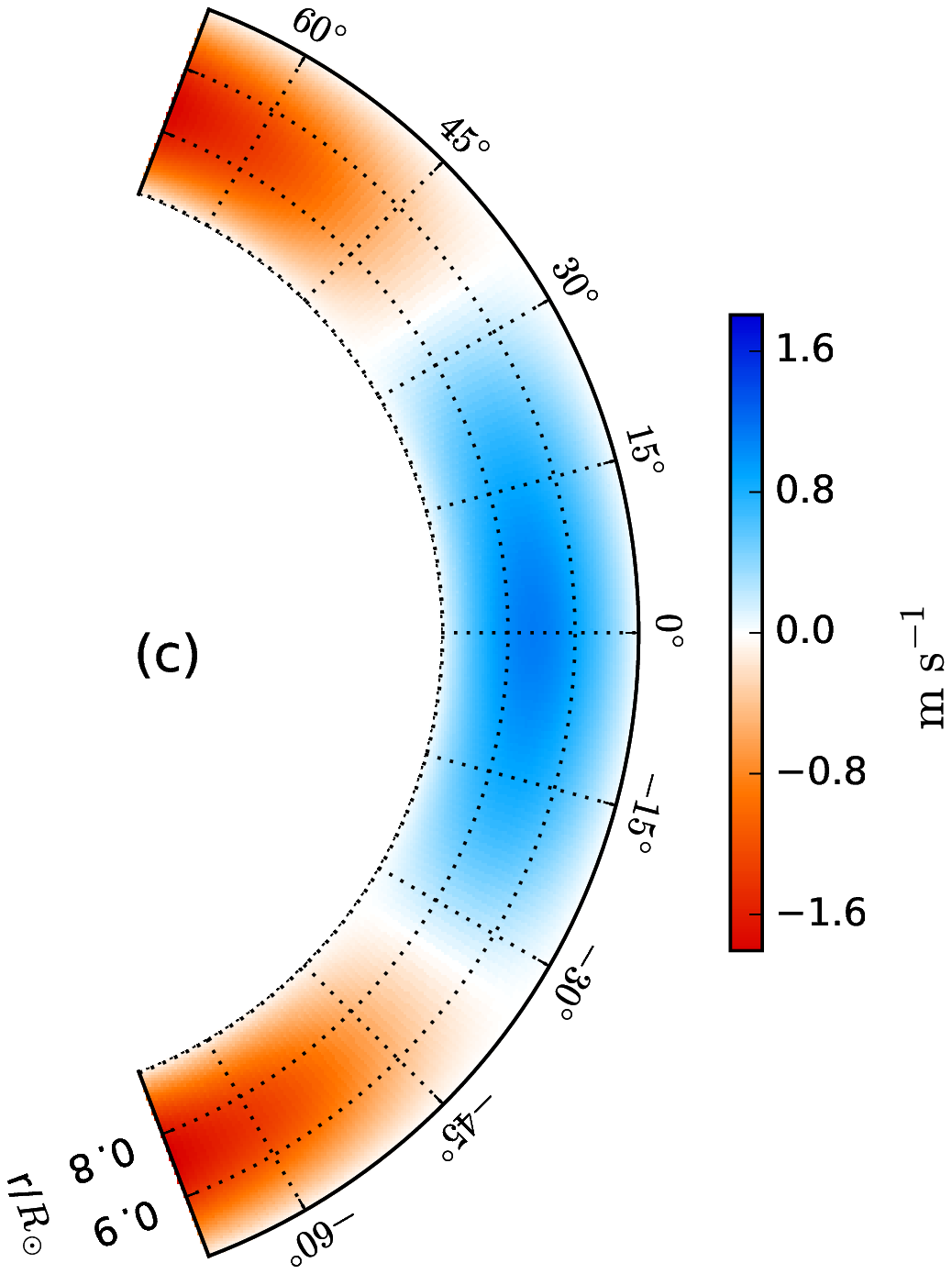}\includegraphics[scale=0.6,clip=true]{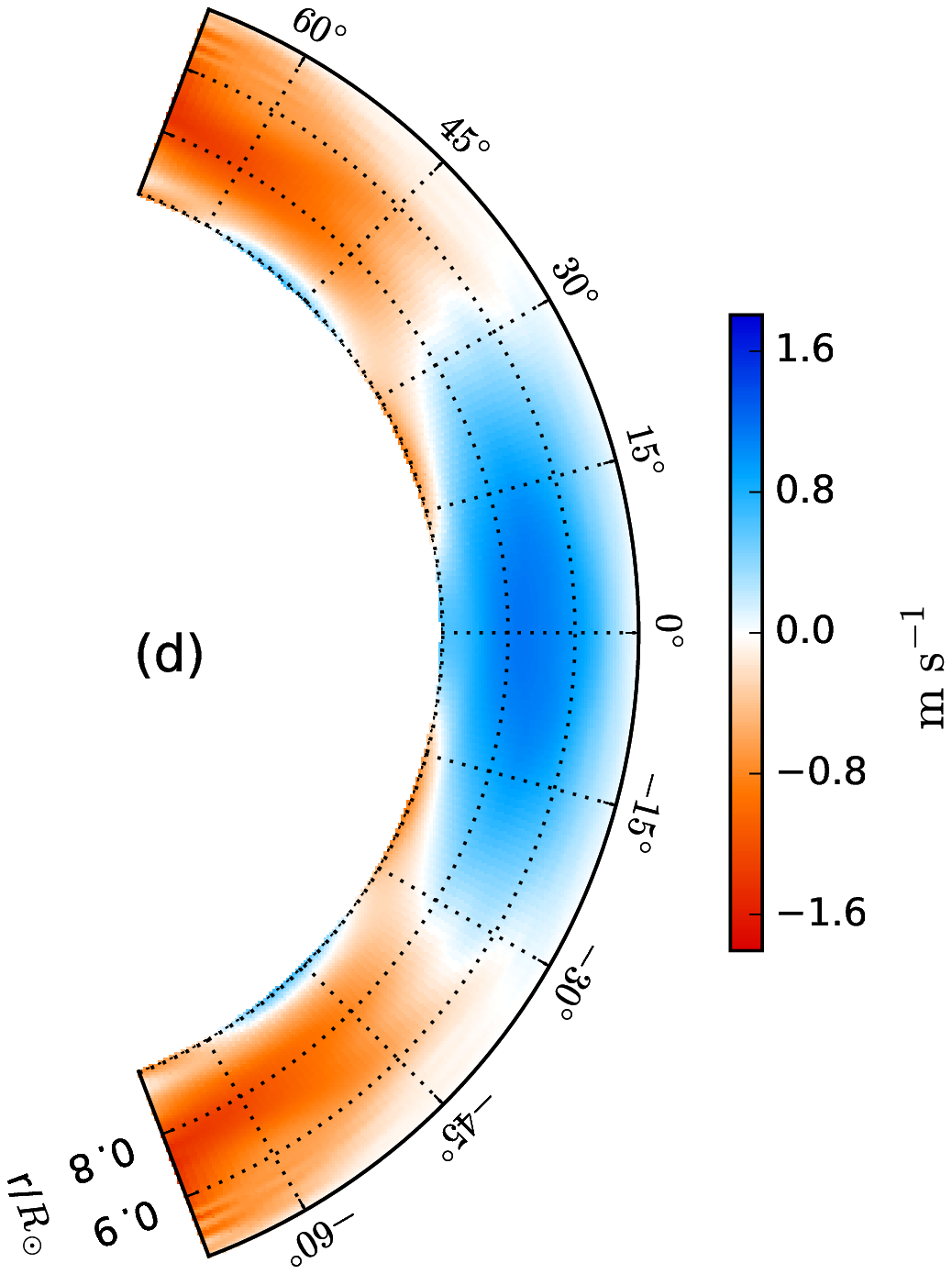} 
\par\end{centering}

\caption{\label{fig:synthetic_single} Synthetic inversion with a single-cell
profile after including noise. We have chosen an input stream-function
profile corresponding to single-cell flow. Panels a and c show profiles
of $v_{\theta}$ and $v_{r}$ respectively obtained from that stream
function using Equation (\ref{eq:streamfn}) and (\ref{eq:streamfn1}).
Panels b and d display the corresponding inverted profile of $v_{\theta}$
and $v_{r}$ using artificial travel times shown in Figure (\ref{fig:synthetic_travel_time}).
Northward velocity is positive and vice versa.}
\end{figure}

\begin{figure}[h]
\begin{centering}
\includegraphics[scale=0.6,clip=true]{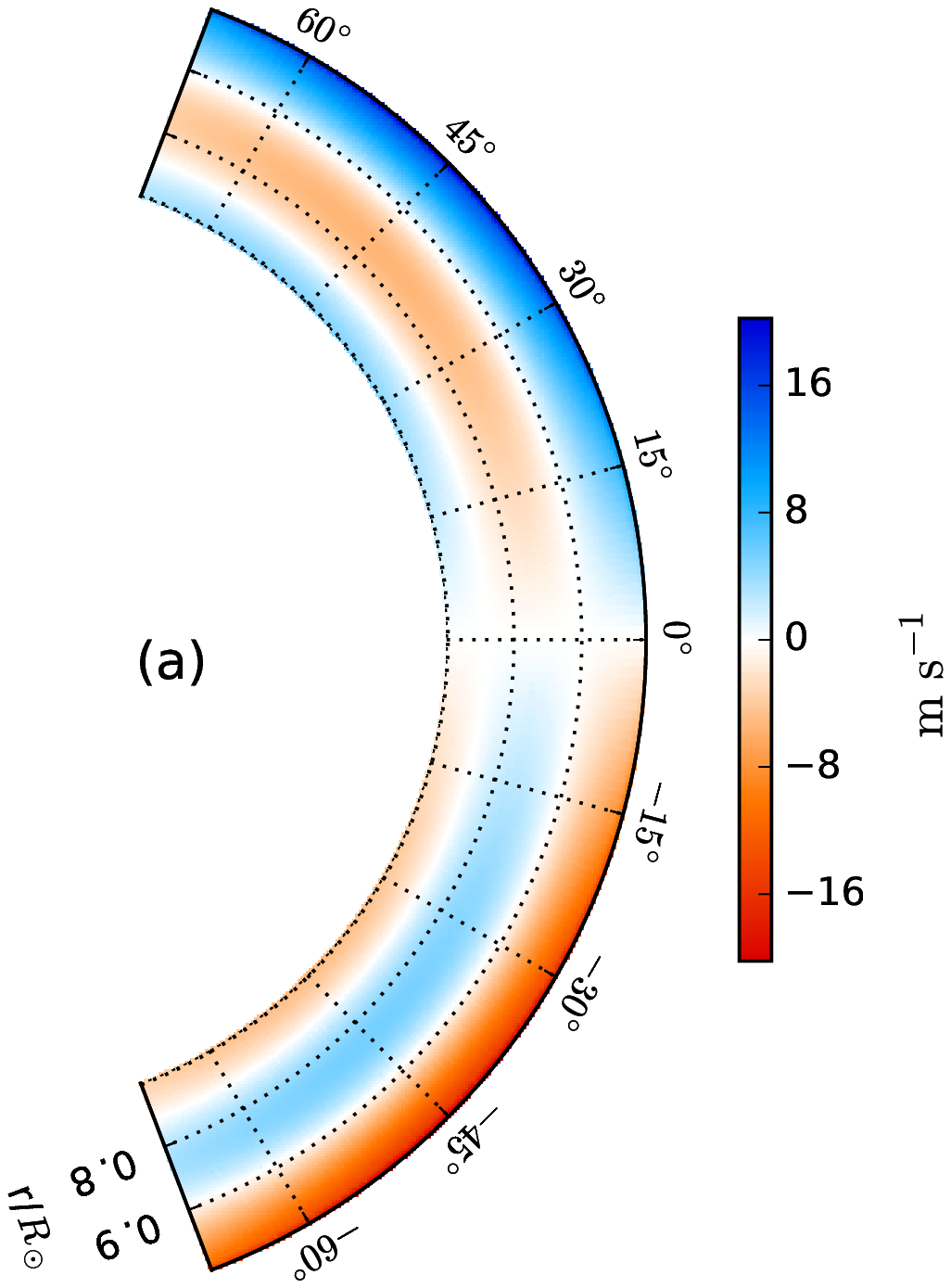}\includegraphics[scale=0.6,clip=true]{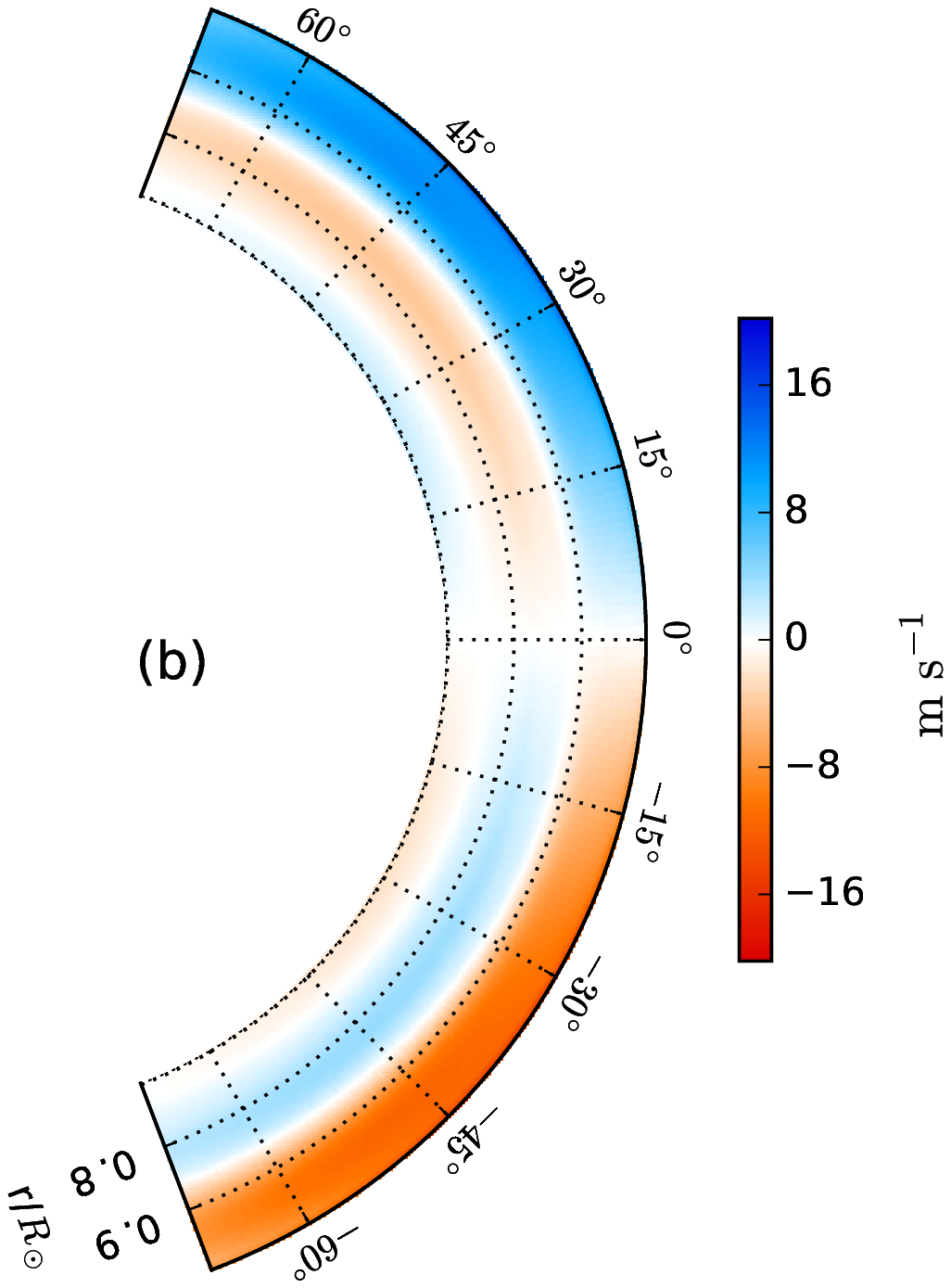} 
\par\end{centering}

\begin{centering}
\includegraphics[scale=0.6,clip=true]{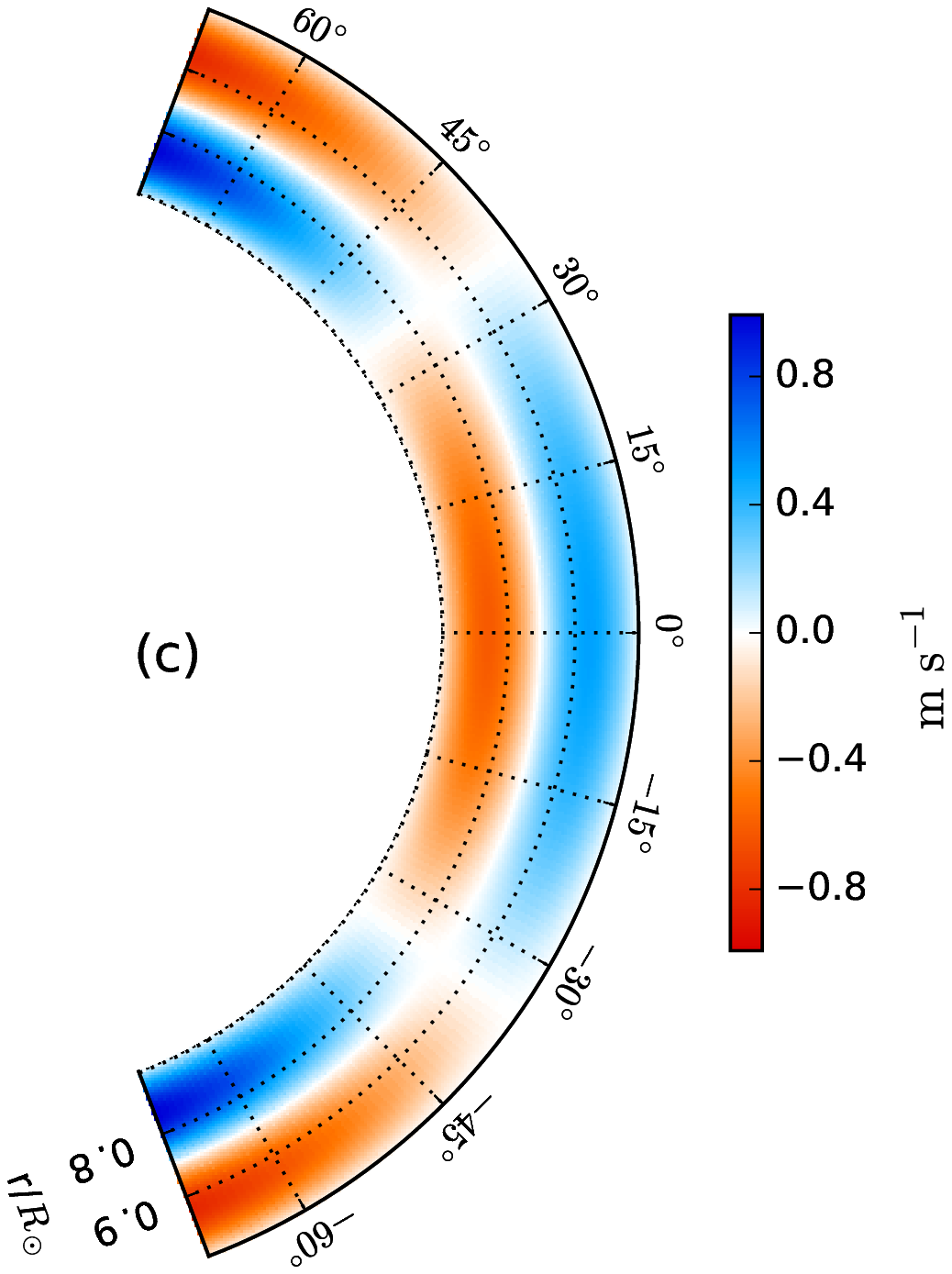}\includegraphics[scale=0.6,clip=true]{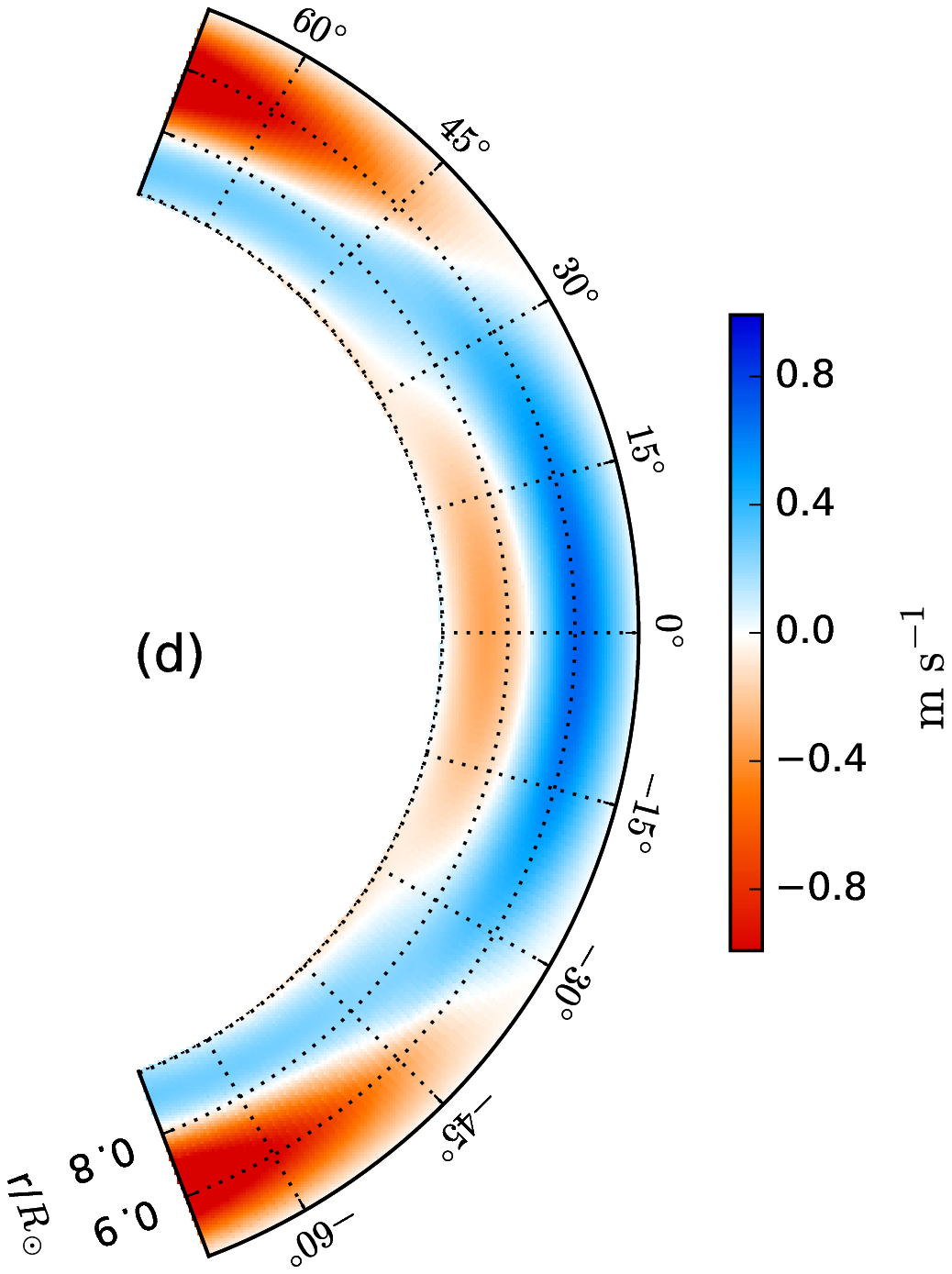} 
\par\end{centering}

\caption{\label{fig:synthetic_double}Synthetic inversion with a double-cell
profile after including noise. Panels a and c show inputs $v_{\theta}$
and $v_{r}$ respectively whereas panels b and d display corresponding
inverted profiles.}
\end{figure}

\begin{figure}[h]
\includegraphics[scale=0.44]{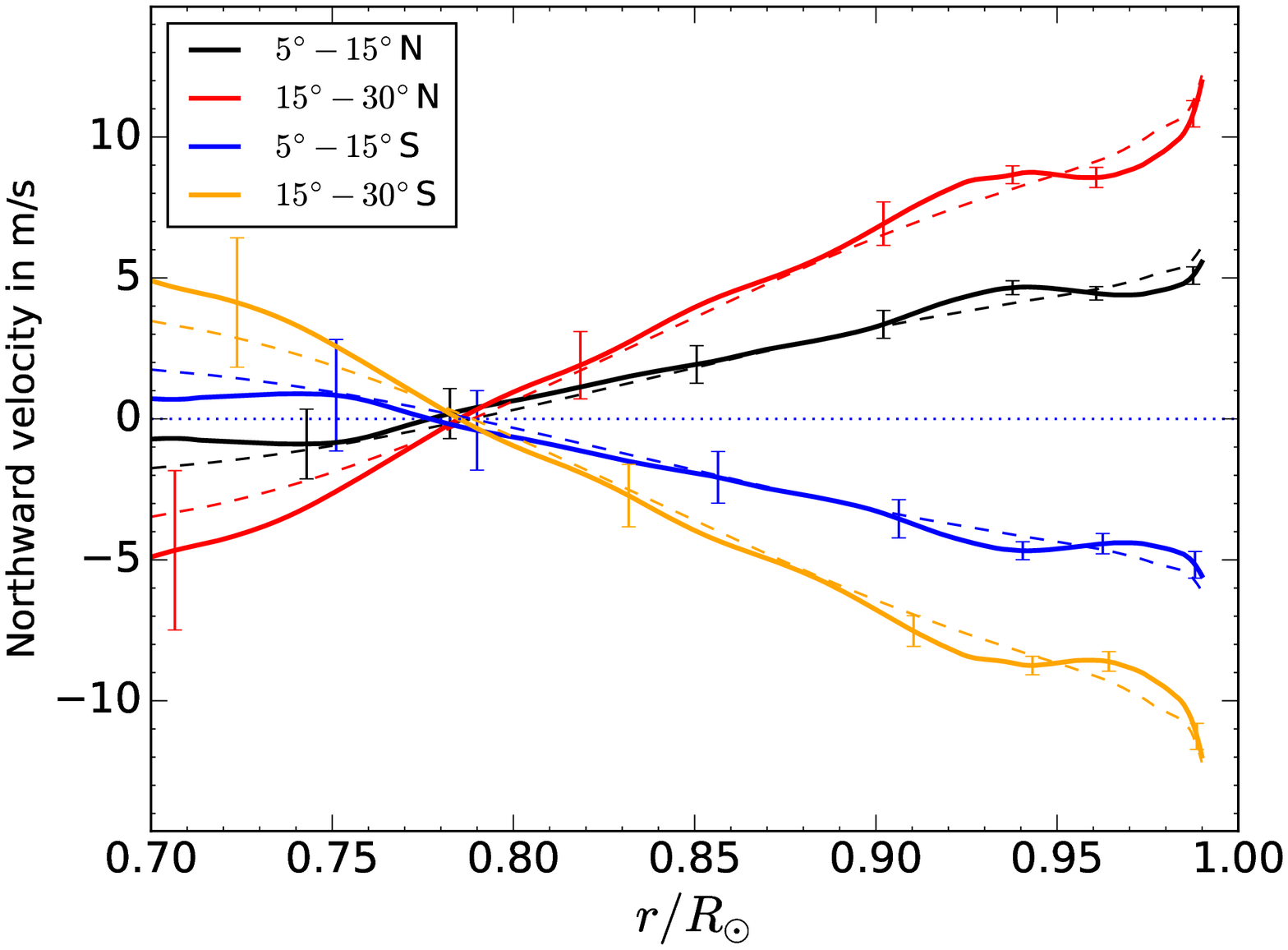}\includegraphics[scale=0.44]{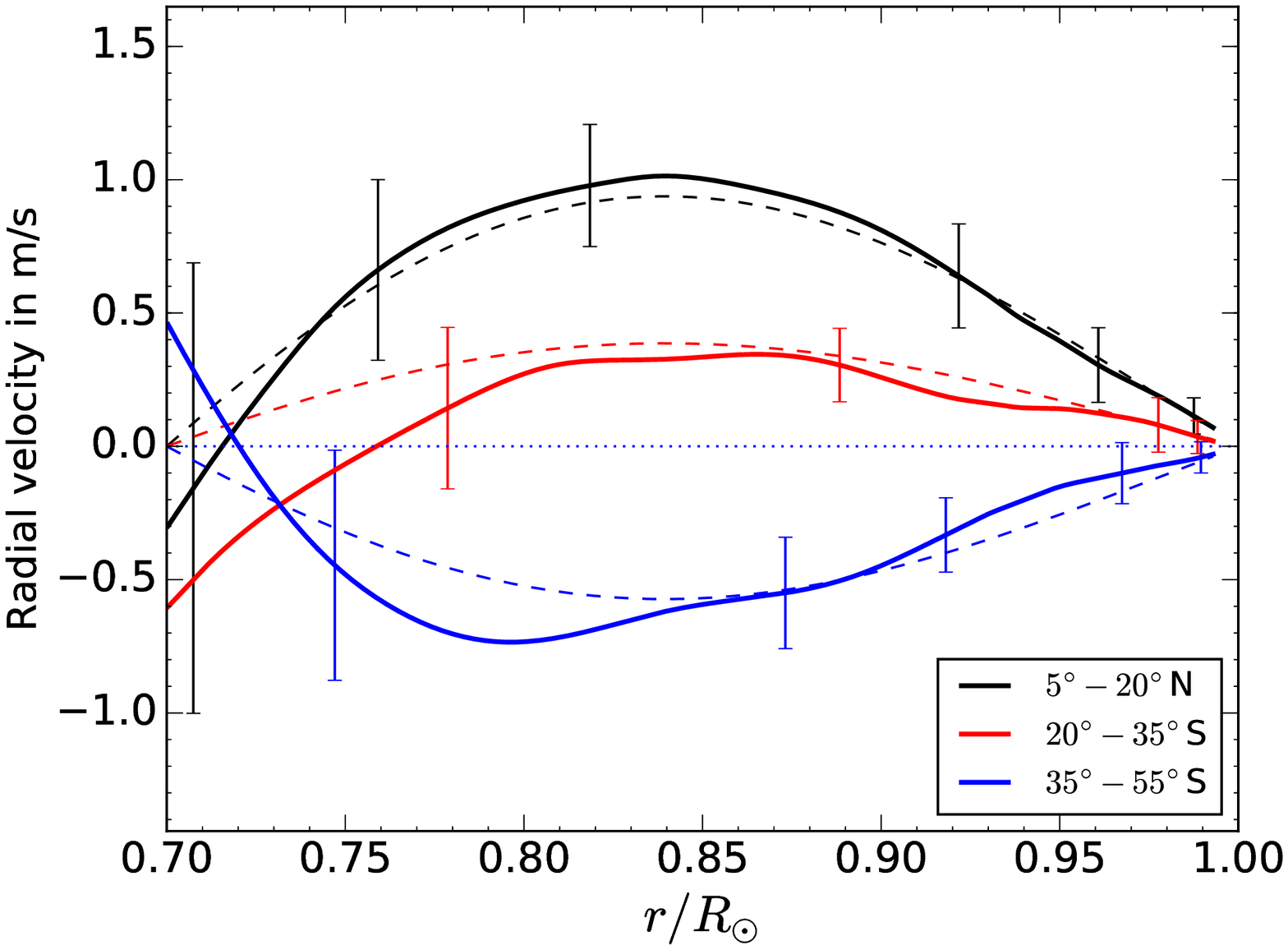}

\includegraphics[scale=0.44]{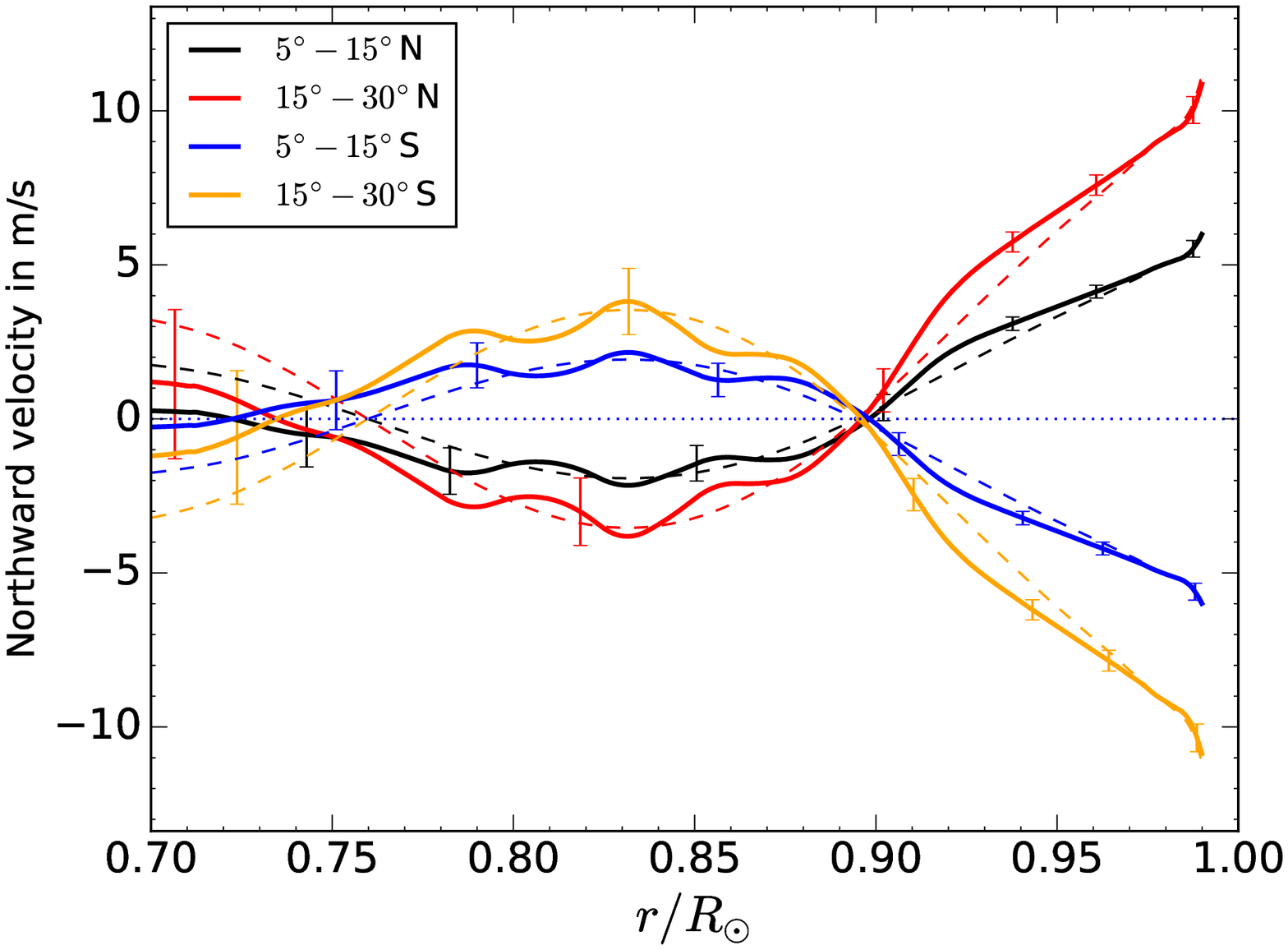}\includegraphics[scale=0.44]{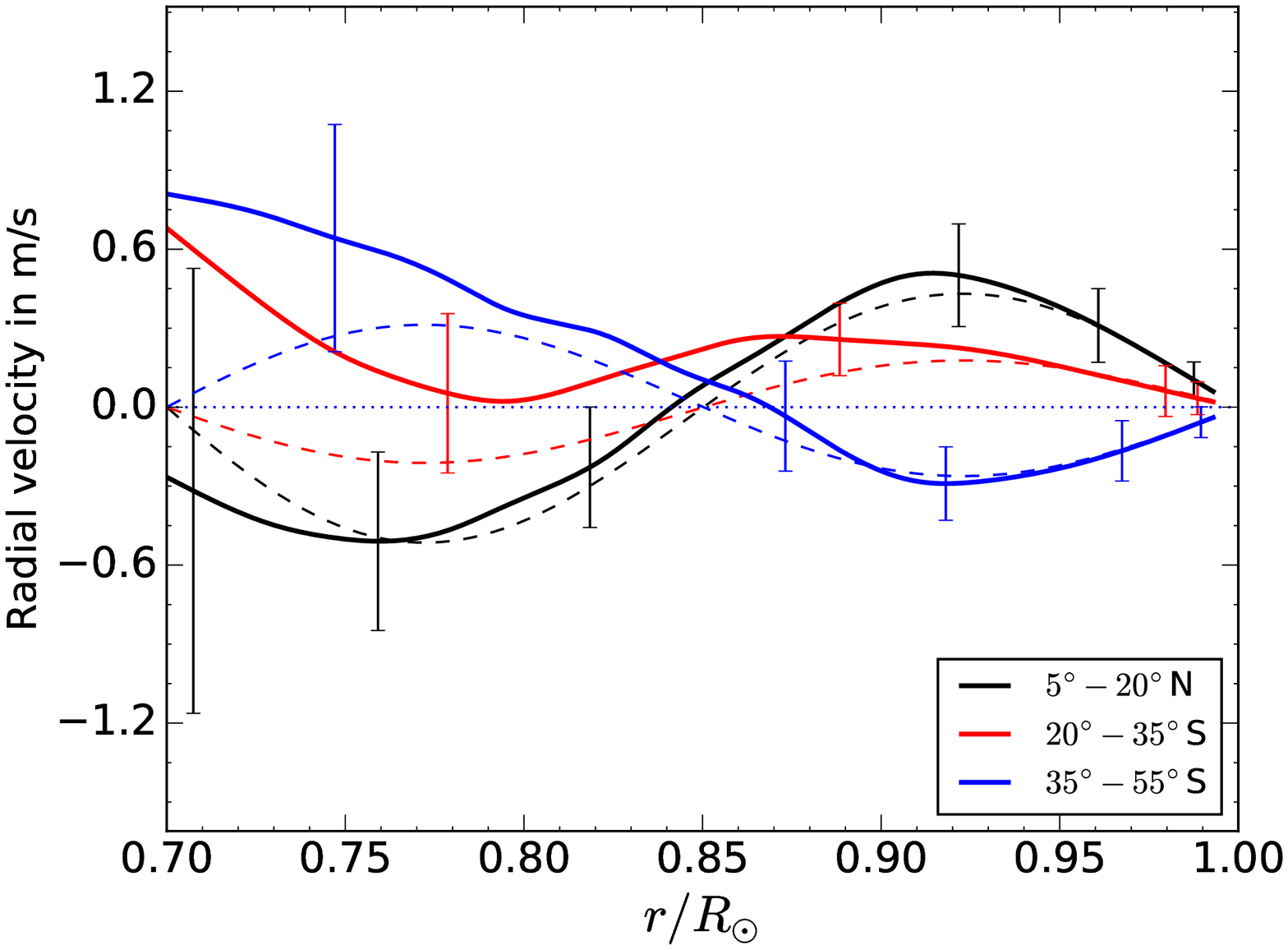}

\caption{\label{fig:synthetic_radial_profile}Synthetic inversion with noise.
Depth dependence of input (dashed line) and inverted profiles (solid
line) of $v_{\theta}$ (left panels) and $v_{r}$ (right panels) averaged
over latitudes mentioned in plot legends are plotted with error bars.
Upper panels and lower panels show results for single-cell case and
double-cell case respectively.}
\end{figure}

We test our inversion algorithm with artificial single and double
cell profiles to verify that we can retrieve both profiles, even in
presence of the observational error. Velocity amplitudes of both the
cells are $20\,\text{m/s}$ at the surface, close to what observationally
found. The single-cell profile extends down to $0.75\,R_{\odot}$
before it changes sign. For the double-cell profile, the first cell
extends down to $0.85\,R_{\odot}$ and second ends at $0.70\,R_{\odot}$.
We generate artificial travel-time data with these two cells and add
random errors. We then choose smoothing parameters in order to obtain
a smooth solution from noisy travel-time data. The results for single
cell profile are shown in Figure (\ref{fig:synthetic_single}). We
have shown both input and inverted profiles together for comparison.
It is seen that our inversion is able to recover the input profile
fairly well. The result for the double-cell profile is shown in Figure
(\ref{fig:synthetic_double}). Again, we find good agreement between
inverted and input profiles. In both cases, we are also able to recover
the radial component of the velocity, $v_{r}$ well. The depth dependence
of the velocity profiles $v_{\theta}$ and $v_{r}$ averaged over
latitude have been shown in Figure (\ref{fig:synthetic_radial_profile})
for single and double-cell cases. We can see that our inversion accurately
recovers the depth profile of $v_{\theta}$. In Figure (\ref{fig:synthetic_travel_time}),
we have compared the input travel-time with that obtained from the
inverted profile by forward modeling. All these results give us the
confidence to proceed further and perform inversions with observed
travel-time measurement data. 

\section{Results}

\begin{figure}[h]
\begin{centering}
\includegraphics[scale=0.6,clip=true]{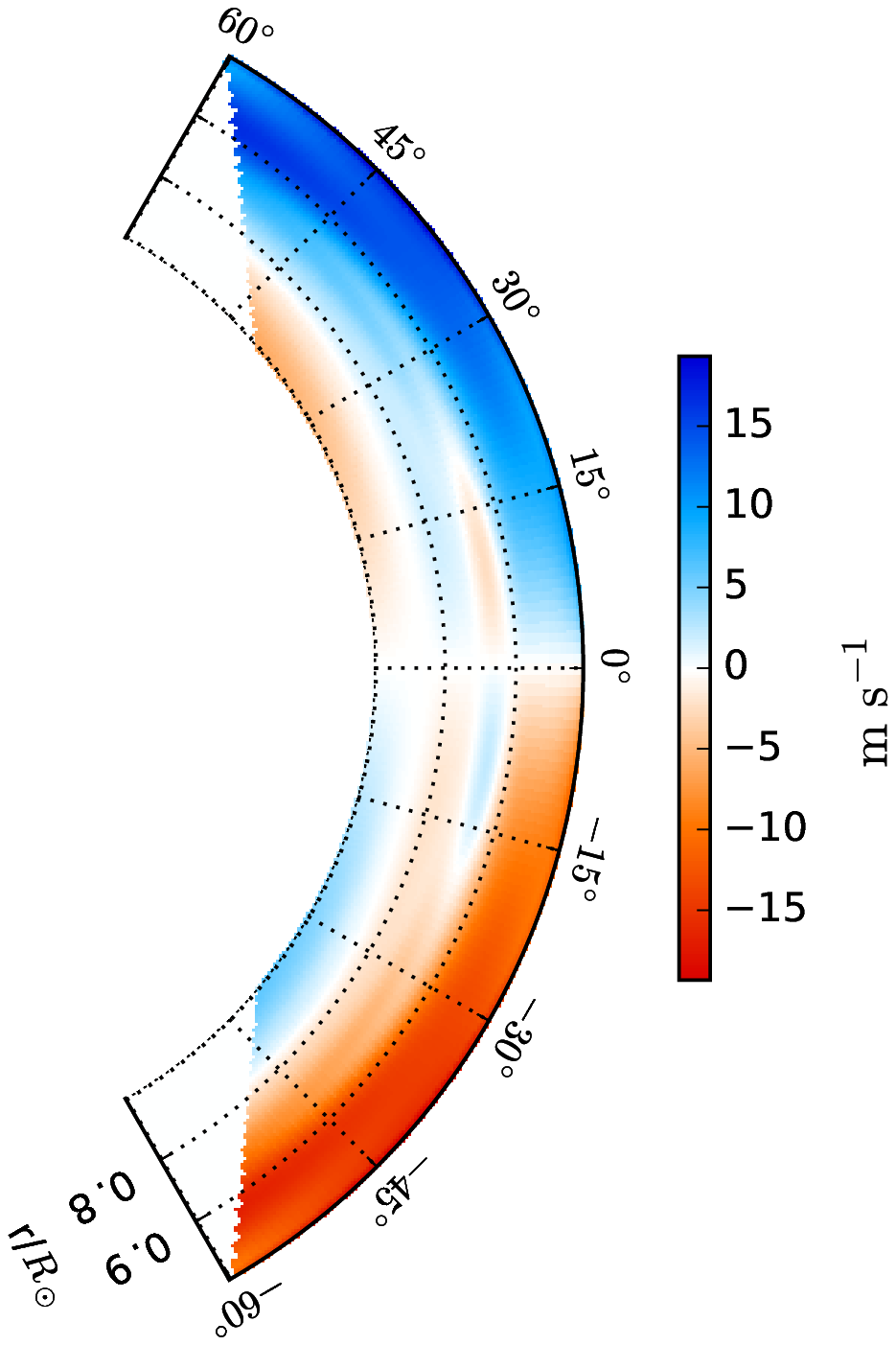}\includegraphics[scale=0.6,clip=true]{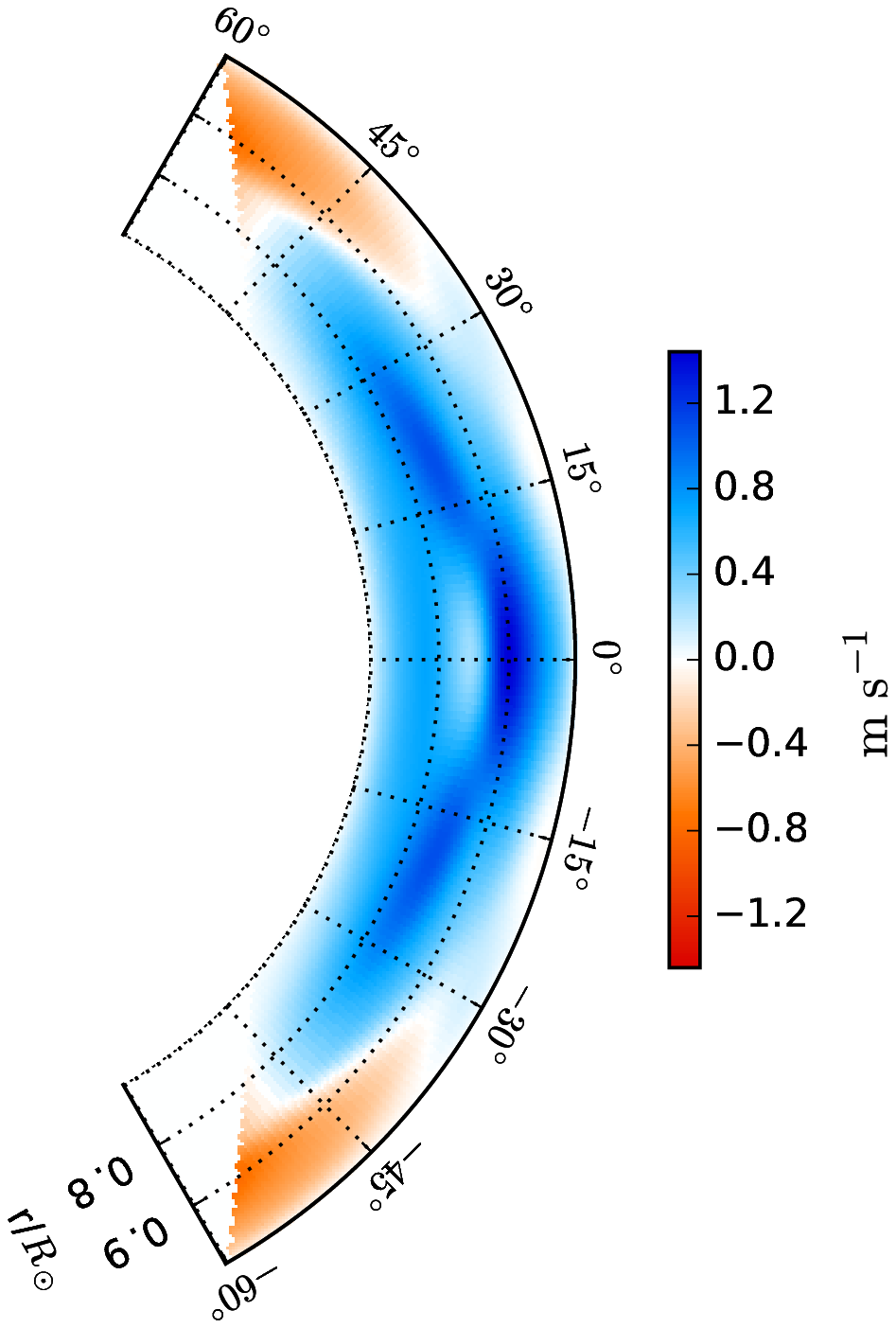} 
\par\end{centering}

\caption{\label{fig:MC_6y_vtheta}Meridional circulation profile after inverting
observed wave travel times. Left panel shows a cross-sectional view
of the horizontal component of the flow $v_{\theta}$ and the right
panel shows the radial component of the flow $v_{r}$ in the $(r,\theta)$
plane. }
\end{figure}

\begin{figure}[h]
\includegraphics[scale=0.45]{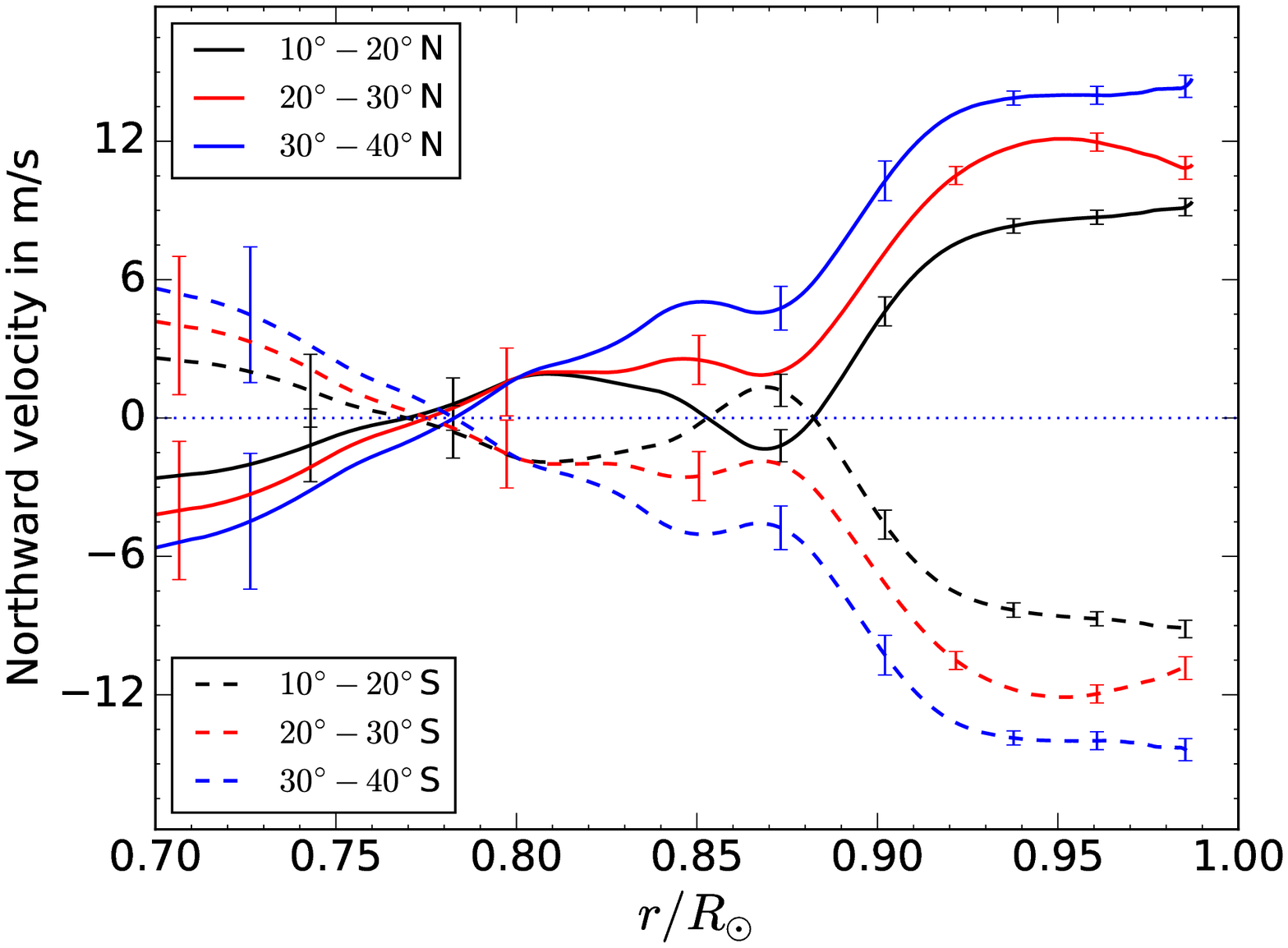}\includegraphics[scale=0.45]{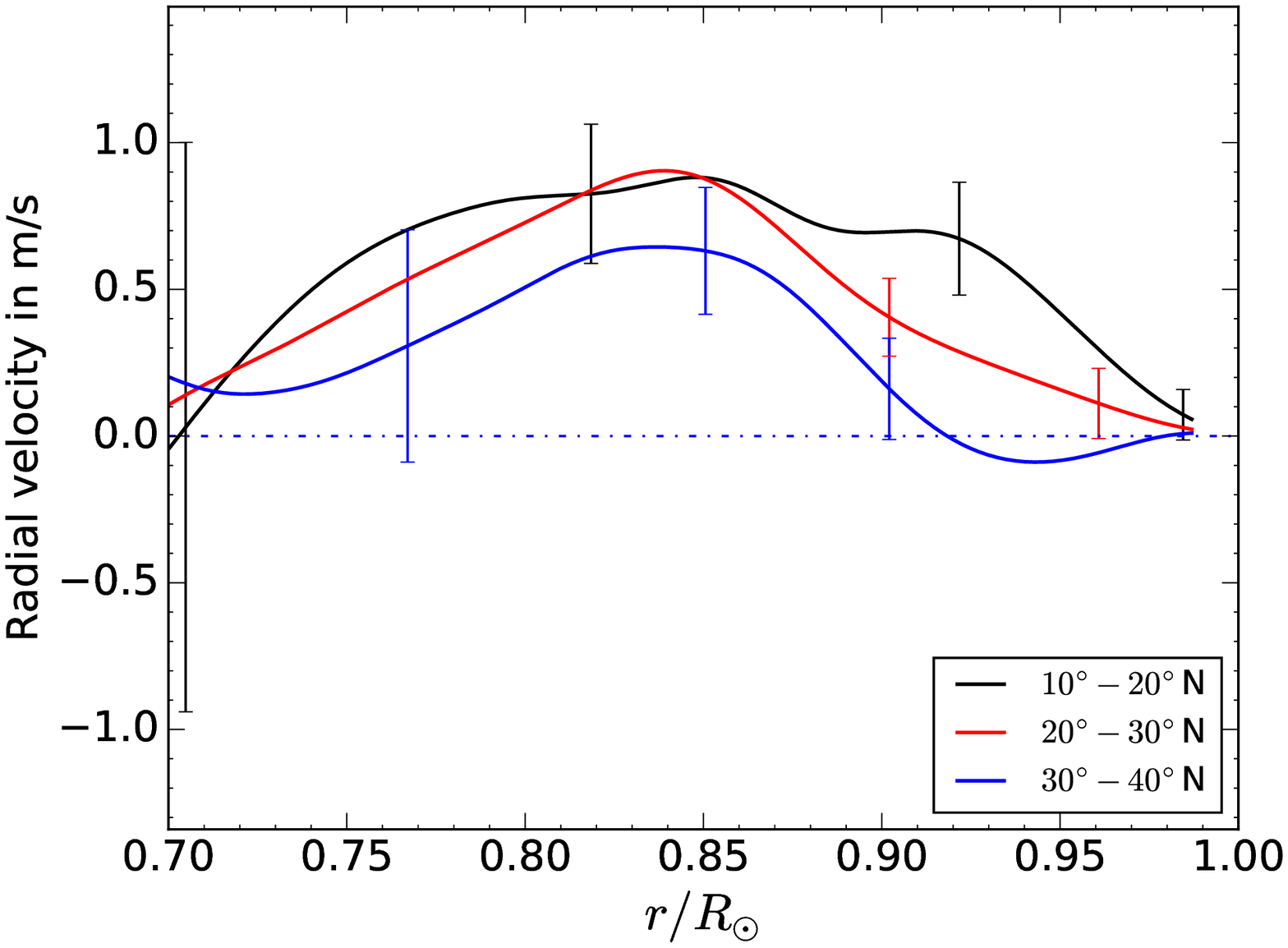}

\caption{\label{fig:cut_v_theta}Meridional circulation profile. Radial dependencies
of $v_{\theta}$ (left panel) and $v_{r}$ (right panel), averaged
over the described latitude range with error bars. }
\end{figure}

We use travel-time measurements of meridional circulation obtained
from analyses of 6 years of observational data by SDO/HMI in the periods
between 2010 to 2016. As mentioned before, we perform inversions to
determine stream function $\psi$ and evaluate components of flow,
$v_{r}$ and $v_{\theta}$, using Equation (\ref{eq:streamfn1}) and
(\ref{eq:streamfn}). We find similar results by using two methods
as described in section (\ref{sec:inversion-technique}). Inversion
results are shown in Figure (\ref{fig:MC_6y_vtheta}) and (\ref{fig:cut_v_theta}).
The results are qualitatively similar to those obtained by \citet{2015ApJ...813..114R}
using ray theory. We plot depth variations of horizontal and radial
flows, $v_{\theta}$ and $v_{r}$, averaged over the latitude range
described in the Figure (\ref{fig:cut_v_theta}). We find that the
return flow is below $0.78\,R_{\odot}$ at all latitudes. Similar to \citet{2015ApJ...813..114R},
we also see sign change in $v_{\theta}$ at low latitudes below $0.9\,R_{\odot}$.
However this is not significant when considering the size of the error
bar (see Figure \ref{fig:cut_v_theta}). We are unable to completely
rule out the picture of multiple-cell meridional circulation because
of this. At higher latitudes, it clearly represents a single-cell
profile. Inverted profile for radial velocity $v_{r}$ (right panel
of Figure \ref{fig:MC_6y_vtheta}) suggests that flow is directed
outward at lower latitude and inward at higher latitude. 

\section{Discussion and conclusion}

Owing to its significant importance for the understanding of solar
dynamo process, there have been several attempts using different techniques
to infer meridional circulation. Due to systematic errors and associated
small magnitudes, inferred models of this circulation tend to vary
widely from one study to another \citep{2013ApJ...774L..29Z,2015ApJ...805..133J,2015ApJ...813..114R,2017ApJ...845....2B,2017ApJ...849..144C}.
In this work, we introduce a combination of techniques to improve
the accuracy of the inference and the condition number of the inverse
problem: (1) to ensure mass conservation we use stream function, (2)
we compute sensitivity kernels for stream function in the first-order
Born approximation and (3) we project the solution on a cubic B-spline,
derivatives of Legendre-polynomial basis to reduce the number of free
parameters (4) we assume hemispheric symmetry and only consider even
Legendre polynomials. In order to validate this method, we test it
on single and double-cell models of the meridional circulation. Surface-velocity
amplitudes of these profiles are chosen so as to be close to the observational
value. We compute synthetic travel times by integrating Born kernels
against the stream function and add randomly generated noise in proportion
to observational errors.

Upon successful validation, we apply the method to infer meridional
circulation from travel-time measurements obtained from 6 years of
SDO/HMI observational data. We find qualitatively similar results
to those of \citet{2015ApJ...813..114R}. Our inversion results suggest
a single cell profile covering $\pm60^{\circ}$ in latitude and up
to $0.7\,R_{\odot}$ in depth in the radial direction. We find a sign
change in the horizontal component of the velocity, $v_{\theta}$
beneath $0.9\,R_{\odot}$ within $\pm20^{\circ}$ in latitude but
is inconclusive considering the size of the error bar. We find equatorward
return flow below $0.78\,R_{\odot}$. 

Our analysis period almost exactly overlaps with analysis period of
\citet{2017ApJ...849..144C}. Differences in the two inversion results
might be due to following factors. We have not removed surface magnetic
regions in order to obtain travel time induced by solar meridional
flow. \citet{2015ApJ...805..165L} have demonstrated that not removing
surface magnetic regions affects the travel time measurements. We
have accounted for local mass conservation and also employed Born
kernels, whereas \citet{2017ApJ...849..144C} have not considered
mass conservation and used ray kernels. 

\citet{2008ApJ...689L.161B} have shown that in order to infer the
depth profile of a weak signal like the solar meridional flow in the
entire convection zone, many years of data are needed. However, during
this period, the flow might change as a function of phase of solar
activity cycle. In that case, we can only obtain an average meridional
flow profile for the time period of the analysis. Improvements such
as including full covariance matrix as \citet{2017ApJ...845....2B}
have suggested and accounting for line-of-sight effects (latter is
only possible in wave-theoretic approach) is reserved for our future
studies. 

K. M acknowledges the financial support provided by the Department
of Atomic Energy, India. SMH acknowledges support from Ramanujan Fellowship
SB/S2/RJN-73 and the Max-Planck Partner group program.

\bibliographystyle{apj}
\bibliography{krishnendu_2nd_proj,krishnendu_bib_Antia}

\end{document}